\documentclass[floatfix, preprintnumbers, amsmath, amssymb, 10pt]{revtex4}
\usepackage{graphicx}
\usepackage{dcolumn}
\usepackage{bm}

\begin{document}

\date{05/30/04}

\title{D0 Matrix Mechanics: New Fuzzy solutions at Large N}

\author{Subodh P Patil}
\email[Electronic address: ]{patil@het.brown.edu}

\affiliation{Dept.of Physics, Brown University,\\
Providence R.I. 02912, U.S.A.\\
{\it and}\\
Dept.of Theoretical Physics,
Tata Institute of Fundamental Research,\\
Homi Bhabha Road, Mumbai 400-005,\\
India}

\begin{abstract}

We wish to consider in this report the large N limit of a particular matrix 
model introduced by Myers describing D-brane physics in the presence of an 
RR flux background. At finite N, fuzzy spheres appear naturally as 
non-trivial solutions to this matrix model and have been extensively 
studied. In this report, we wish to demonstrate several new classes of solutions which appear in the large N limit, corresponding to the fuzzy cylinder,the fuzzy plane and a warped fuzzy plane. The latter two solutions arise from a possible "central extension" to our model that arises after we account for non-trivial issues involved in the large N limit. As is the case for finite N, these new solutions are to be interpreted as constituent D0-branes forming D2 bound states describing new fuzzy geometries.
\end{abstract}

\maketitle

\begin{flushleft}

\section{introduction}
Non commutative geometry, not withstanding its intrinsic interest 
\cite{k1}\cite{k2}\cite{k3} has attracted much attention recently through 
its apearance in string theory and M-theory  (see 
\cite{k4}\cite{k5}\cite{k6}\cite{k7} and references therein). One of the 
more familiar manifestations of non commutative geometry in this context 
arises when one considers the dynamics of D-branes \cite{k8} in various 
backgrounds. There is a natural matrix model description of this dynamics, 
and in general one expects the solutions of this matrix model to correspond 
to various fuzzy geometries. In our report, we wish to study a particular 
matrix model that arises when we consider the dynamics of D0 branes in the 
presence of an RR flux background \cite{k5}\cite{cs1}. Although this model arises in 
various other contexts (see \cite{k4}), we will only be interested in this 
model in so far as it provides for a framework in which to study the 
dynamics of noncommutative spaces. The interpretation we prefer is that 
which arises from Myers' \cite{k5} derivation of this matrix model: that the 
dynamics and energetics of this model correspond to various bound D0 brane 
configurations. The dimension of the matrices we consider corresponds to the 
number of D0 branes we have in our system, and so the large N limit 
corresponds to taking the limit of a large number of D0 branes. This is 
certainly a reasonable scenario to consider from the perspective of early 
universe physics. Indeed it is with with one eye on spacetime physics that 
we undertake this study, as we wish to use this report as a point of 
departure for future work \cite{k9}, where there are preliminary indications 
of dynamical D-brane topology change in this large N limit. \linebreak

\par

Although it might seem at first sight that the large N limit involves a loss 
of calculational ease when considering the dynamics of the system, we find 
that after careful consideration the situation is not that much different 
from the case at finite N. The few important diffrences we will uncover only 
serve to enrich the problem, and allow us solutions that we would not have 
had at finite N. In particular, we find that such considerations yield 
solutions that correspond to the D0-branes configured as non-commuting 
planes, and non-commuting cylinders. The possibility of transitions between 
these new geometries will be taken up in a follow up report \cite{k9}.
\section{The model: a first pass}

The matrix model we will be studying is described by the following action 
(\cite{k4}\cite{k5}\cite{k6}\cite{k7}):

\begin{equation}
\label{act}
S = T_0 \int dt Tr \Bigl( \frac{1}{2} \dot{X}_i \dot{X}_i + 
\frac{1}{4}[X_i,X_j][X_i,X_j] - \frac{i}{3}\kappa \epsilon_{ijk}X_i[X_j,X_k] 
\Bigr)
\end{equation}

Following the conventions of \cite{k4}, we work in units where $2\pi\alpha' 
= 1$. The $X_i$ are taken to be $N \times N$ traceless, hermitian matrices and $T_0 = 
\sqrt{2\pi}/g_s$ is the D0 brane tension. The index $i$ runs from 1 to 3. In 
addition to this action, we have the Gauss law constraint:

\begin{equation}
\label{Gauss}
[\dot{X}_i,X_i] = 0
\end{equation}

Which arises from the $A_0$ equation of motion for the RR gauge field. Note that the last term in the action is a Chern-Simons term which induces interactions between the D0-branes through the 4-form flux, which assumes the vacuum expectation value $F^{4}_{0123} = -2\kappa\epsilon_{ijk}$. This term was deduced by Myers by demanding consistency of the D-brane action with T-duality \cite{k5}.\linebreak

\par
Let us for the moment forget dynamics, and concentrate on the static 
solutions of this model. Varying the action gives us the static equation of 
motion:

\begin{equation}
\label{stateom}
[X_j, \bigl([X_i,X_j] - i\kappa \epsilon_{ijk}X_k \bigr)] = 0
\end{equation}

From which we can immediately deduce two classes of solutions: commuting 
matrices ($[X_i,X_j] = 0$) and fuzzy two spheres-

\begin{equation}
\label{fuzzysp}
[X_i,X_j] = i\kappa \epsilon_{ijk}X_k
\end{equation}

Any representation of the lie algebra of SU(2), irreducible or otherwise, 
given by the generators $\{J_i\}$ furnishes the solution (see \cite{k10} for 
a review of the non commutative 2-sphere):

\begin{equation}
\label{sp}
X_i = \kappa J_i
\end{equation}

Where for an irrep of su(2), the casimir operator $X_i X_i = \kappa^2 
j(j+1)$ gives us the radius squared of the 2-sphere and the dimension of the 
irrep is given by $N = 2j+1$. Clearly, a reducible representation can be 
expressed as a direct sum of irreps:

\begin{equation}
\label{rep}
X_i = \oplus_{r} \kappa J_i^{(r)}
\end{equation}

The summation over $r$ runs over the irreps included in this representation 
and describes a solution of multiply superimposed fuzzy 2-spheres. As 
before, the radius of each constituent fuzzy 2-sphere is given by $R^2_r = 
\kappa^2 j_r(j_r+1)$ and the dimension of the representation is now given $N 
= \sum_r (2j_r+1)$.

The energy of these solutions is given by their potential energy:

\begin{equation}
\label{V}
V = Tr \Bigl( -\frac{1}{4}[X_i,X_j][X_i,X_j] + \frac{i}{3}\kappa 
\epsilon_{ijk}X_i[X_j,X_k] \Bigr)
\end{equation}

From which we see that commuting matrices correspond to zero energy 
solutions. Without the presence of the Chern-Simons term in our model, these 
solutions would span the lowest energy configurations. However the addition 
of the Chern-Simons term modifies things drastically by permitting fuzzy 
sphere solutions, which have a negative energy:

\begin{equation}
\label{esph}
E = -T_0\kappa^4\frac{1}{6}\sum_r j_r(j_r+1)(2j_r+1)
\end{equation}

So we see that the fuzzy two spheres describe a bound state of D0 branes, 
with the irreducible representations corresponding to the lowest energy 
configuration for a given N. The observation that reducible representations 
have greater energy than irreducible ones has motivated the belief that this 
model could describe topology changing physics through transitions between 
these states. The simplest example of a transition that this model might be 
capable of would be when two separate spheres meld into one. Observations 
based on studying the energetics of this model \cite{k4}\cite{k6}\cite{k7} 
have yielded promising preliminary results, but an explicit solution 
interpolating between two topologies is still lacking. We wish to use the 
results collected in this paper to further investigate the existence of 
topological transtions in a future report \cite{k9}, and in the process 
hopefuly highlight why these solutions cannot be taken for granted even if 
the energetics of the model strongly imply the possibility of topology 
change.\linebreak

\par

Returning to the static equations of motion (\ref{stateom}), it might appear 
that we have exhausted all the possible static solutions, however a slight 
recasting of the problem will show that this is not the case. Performing the 
change of variable:

\begin{equation}
X_+ = X_1 + iX_2 ~,~ X_- = X_1 - iX_2~;~ X_+^\dag = X_-
\end{equation}

The equations of motion become:

\begin{equation}
\label{lc1}
\ddot{X}_+ ~= ~\frac{1}{2}[X_+, [X_+,X_-] ] ~+ ~[X_3, [X_+,X_3] + 2\kappa 
X_+ ]
\end{equation}

\begin{equation}
\label{lc2}
\ddot{X}_- ~= ~\frac{1}{2}[X_-, [X_-,X_+] ] ~+ ~[X_3, [X_-,X_3] - 2\kappa 
X_- ]
\end{equation}

\begin{equation}
\label{lc3}
\ddot{X}_3 ~= ~\frac{1}{2}[X_+, [X_3,X_-] ] ~+ ~\frac{1}{2}[X_-, [X_3,X_+] ] 
~+ ~\kappa [X_+,X_-]
\end{equation}

And the Gauss law condition becomes:

\begin{equation}
\label{gc}
[\dot{X}_3,X_3] ~+ ~\frac{1}{2}[\dot{X}_+,X_-] ~+ 
~\frac{1}{2}[X_+,\dot{X}_-] ~= ~0
\end{equation}

Since (\ref{lc1}) and (\ref{lc2}) are adjoints of each other, we are left 
with only two independent matrix equations to satisfy. The fact that one of 
the equations governs the motion of a non-hermitian matrix, which in general 
has a hermitian and an anti-hermitian part, accounts for the two hermitian 
equations we had previously. We now make the ansatz:

\begin{equation}
\label{ansatz}
[X_+,X_-] = \lambda X_3 ~,~ [X_3,X_\pm] = \pm \theta X_\pm
\end{equation}

For which eqs. (\ref{lc1})-(\ref{lc3}) imply

\begin{eqnarray}
\label{cond}
\ddot{X}_+ &=& (2\kappa - \theta - \lambda /2)\theta X_+\\ \ddot{X}_3 &=& 
(\kappa - \theta)\lambda X_3
\end{eqnarray}

Thus in the static case, we can immediately read off three distinct 
solutions. $\lambda = \theta = 0$ is one solution which corresponds to 
commuting matrices. $\lambda = 2\kappa$, $\theta = \kappa$ is another, which 
corresponds to the already known fuzzy two sphere solutions. It should be 
clear that the change of variable we made is identical to constructing the 
usual raising and lowering operators out of $X_1$ and $X_2$, and the 
algebra:

\begin{equation}
\label{su2rl}
[X_+,X_-] = 2\kappa X_3 ~,~ [X_3,X_\pm] = \pm \kappa X_\pm
\end{equation}

is none other than that of su(2). The third solution is given by $\lambda = 
0$, $\theta = 2\kappa$:

\begin{equation}
\label{cyl}
[X_+,X_-] = 0 ~,~ [X_3,X_\pm] = \pm 2\kappa X_\pm
\end{equation}

Which corresponds to the fuzzy cylinder \cite{k11} provided we impose the 
additional constraint $X_+X_- = \rho^2 I$. Since the non-commutative 
cylinder is not as familiar a fuzzy geometry as the sphere or the plane, we 
briefly outline how it is constructed, following the treatment of 
\cite{k11}. We wish to interject at this point that alarm bells should 
already be ringing, as (\ref{cyl}) describes the lie algebra of a non 
semi-simple group (re-expressing the above in terms of $X_1, X_2, X_3$ we 
see that $X_1$ and $X_2$ form a proper ideal). Hence we will neccesarily be 
dealing with infinite dimensional representations if we want to preserve the 
hermitian nature of the generators (i.e. if we want to work in a unitary 
representation)-- which is required by the very nature of our model. It is 
not immediately clear what equations of motion the action (\ref{act}) will 
yield as $N$ tends to infinity, as crucial in our derivation of the 
equations of motion was the cyclic property of the trace, which allowed us 
to bring all of the variations of the $X_i$ to one side of the 
trace\footnote{In general the variations will not commute with the matrices 
themselves and for this reason the cyclic property of the trace proves 
essential in applying the variational principle.}. The cyclic property of 
the trace cannot be taken forgranted in infinite dimensions, and for 
unbounded operators it generally fails. We shall see in the next section 
that rather remarkably, we can be forgiven for proceeding as we are at the 
minute. Indeed, there we will discover a possible fourth class of solutions after 
accounting for the subtleties of the infinite N limit. However for the 
present purposes we will take these issues forgranted and justify our 
treatment here later.\linebreak

\par

Returning to the fuzzy cylinder, it helps if we begin with the commutative 
cylinder, described by a non compact coordinate $\tau ~ \epsilon ~ \mathbb 
R$ and an angular coordinate $\phi ~ \epsilon ~ [0,2\pi]$. We see that 
functions on the cylinder are spanned by the basis:

\begin{equation}
\label{basis}
\{ e^{in\phi} \}_{n \epsilon \mathbb Z} ~;~ f(\tau,\phi) = \sum_n c_n(\tau) 
e^{in\phi}
\end{equation}

Defining $x_+ = \rho e^{i\phi}$, $x_- = \rho e^{-i\phi}$, where $\rho$ is 
the radius of the cylinder, we can generate all functions on the cylinder by 
a power series in these variables. Furthermore, we have the relation $x_+x_- 
= \rho^2$. The Poisson structure of the cylinder is defined by the brackets:

\begin{equation}
\label{pb}
\{f,g\} := \frac{\partial f}{\partial \tau}\frac{\partial g}{\partial \phi} 
- \frac{\partial g}{\partial \tau}\frac{\partial f}{\partial \phi}
\end{equation}

where $f$ and $g$ are arbitrary functions on the cylinder. By the Leibniz 
rule, all Poisson brackets can be generated from the following elementary 
brackets:

\begin{equation}
\label{fund}
\{\tau,x_\pm\} = \pm i x_\pm ~,~ \{x_+,x_-\} = 0
\end{equation}

The relationship $x_+x_- = \rho^2$ is preserved by these brackets (i.e. it 
is central):

\begin{equation}
\label{cent}
\{\tau,x_+x_-\} = \{x_\pm,x_+x_-\} = 0
\end{equation}

Derivatives of functions on the cylinder can now be effected by the action 
of the brackets:

\begin{equation}
\label{derv}
\partial^2_\phi f = \{\tau,\{\tau,f\}\} ~;~ \partial^2_\tau f = 
\frac{1}{\rho^2}\{x_-,\{x_+,f\}\}
\end{equation}

To obtain the non-commutative cylinder, one `quantizes' the Poisson 
structure through the prescription $\{ , \} \to \frac{i}{\lambda}[ , ]$ 
where $\lambda$ is the non-commutativity parameter. Although one feels that 
this procedure shouldn't raise any suspicion due to its similarity with 
canonical quantization, it is in fact a very well defined and well motivated 
prescription for constructing fuzzy geometries \footnote{ See ref \cite{k12} 
and references therein for an excellenct introduction to the procedure for 
quantizating classical manifolds.}. In fact a quick and easy way of seeing 
how this is plausible, is to realise that because of the remarkable formula 
$[A,BC] = [A,B]C + B[A,C]$, the above quantization prescription reproduces 
all Poisson brackets of functions, and hence their derivatives, in the form 
of commutator brackets acting on the corresponding operator functions, up to 
ordering. Returning to the problem at hand, we can thus write down the 
structure relations of the fuzzy cylinder as:

\begin{equation}
\label{fuzzyc}
[\tau,x_\pm] = \pm \lambda x_\pm ~,~ [x_+,x_-] = 0 ~,~ x_+x_- = \rho^2
\end{equation}

Thus if we take $\lambda = 2\kappa$, we see that the solution to (\ref{cyl}) 
is indeed the fuzzy cylinder if we identify $X_3$ with $\tau$ and $X_\pm$ 
with $x_\pm$. From the algebra (\ref{cyl}) one can immediately glean that 
the spectrum of $X_3$ is integer multiples of $2\kappa$ and the $X_\pm$ act 
as raising and lowering operators in the basis where $X_3$ is diagonal. In 
fact because of the relation $X_+X_- = \rho^2$, we can conclude that the 
action of $X_+$ is that of $\rho$ times the elementary shift operator:

\begin{equation}
\label{shift}
X_3|n\rangle = 2\kappa n |n\rangle ~\to~ X_+|n\rangle = \rho |n+1\rangle
\end{equation}

Where it should be clear that $X_-$ performs the opposite shift. It is this 
qualitative similarity with the situation for the 2-sphere (where of course 
the raising operator has a more complicated dependence on the state it is 
acting on through the Clebsch-Gordon coefficients) that motivates us to 
investigate in \cite{k9}, the possibility of a transition between these two 
fuzzy geometries.\linebreak

\par

Thus in addition to forming bound states corresponding to fuzzy 2-spheres, a 
large number of D0 branes in the presence of RR flux can also form a bound 
state corresponding to the fuzzy cylinder. However there is more to this 
model yet, in the next section we will confront the issues involved in 
dealing with infinite dimensional matrix actions, and find through our 
efforts yet another distinct fuzzy geometry admitted by this matrix model-- 
the noncommutative plane.

\section{Infinite dimensional matrix dynamics}

In deriving the equations of motion from the action (\ref{act}) defining 
this matrix model, we had to make use of two crucial aspects of the trace. 
The first aspect is that it is cyclic-- for an arbitrary variation $\delta 
X_i$ will not commute with the $X_i$ themselves, hence we need this 
cyclicity to bring all variations to one side of the trace. The second 
aspect we need is the positive definiteness of the trace norm, which allows 
us to deduce the equations of motion from the statement that the action is 
extremized by all variations in the fields:

\begin{equation}
\label{trnorm}
Tr \Bigl( \frac{\delta L}{\delta X_i} \delta X_i \Bigr) = 0 ~\forall~ \delta 
X_i ~\to~ \frac{\delta L}{\delta X_i} = 0
\end{equation}

Now in the case where the matrices we are dealing with are infinite 
dimensional, and the trace is over a suitable Hilbert space, we cannot take 
the cyclic property of the trace forgranted. Certainly the trace remains 
cyclic for bounded operators, but when we are dealing with unbounded 
operators this is not the case (e.g. $X$, where the spectrum is the real 
line). The most famous example comes from the canonical pair $X$ and $P$ 
with $[X,P] = i$. It should be apparent that if were we to work in a basis 
where any one of the pair is diagonal, then clearly $Tr XP \neq Tr PX$. In 
fact far from being a mysterious mathematical oddity, the origin of the 
non-cyclicity of the trace lies in the mundane fact that one cannot 
interchange the order of integration over a domain where the function we are 
integrating is unbounded.\linebreak

\par

This immediately begs the question of how we should proceed deriving the 
equations of motion if we want to drop the assumption that the matrices we 
are working with are finite dimensional. The answer is just as immediate and 
for the most part, the end result is not that different in many respects 
except for one, which will facilitate a "central extension" to our matrix 
model. Before we continue however, we wish to point out that the only 
regularization procedure we'd ever need to carry out in order to make the 
transition to infinite dimensional matrices, is to incorporate a 
normalization factor into the trace such that $Tr I = 1$. This is only 
neccesary when studying the energetics of the system, and has no physical 
significance in terms of a cut-off length scale. Rather, all this serves to 
do is to factor out the divergent behaviour of the energy such that we can 
meaningfully compare the energetics of different configurations through 
their energy densities, which always remains finite.\linebreak

\par

Varying the action (\ref{act}) as it stands, we end up with the following:

\begin{eqnarray}
\label{var}
\delta S &=& T_0 \int dt Tr \Bigl( \frac{1}{2} \delta\dot{X}_i\dot{X}_i + 
\frac{1}{2} \dot{X}_i\delta\dot{X}_i + \frac{1}{2}[\delta X_i, X_j][X_i,X_j] 
+ \frac{1}{2}[ X_i, X_j][\delta X_i,X_j]\\ &-& 
\frac{i\kappa}{3}\epsilon_{ijk}\delta X_i[X_j,X_k] - 
\frac{i2\kappa}{3}\epsilon_{ijk}X_i[\delta X_j,X_k] \Bigr)
\end{eqnarray}

Where for finite dimensional matrices, we compensated for the 
non-commutativity of the fields and their variations by using the cyclic 
property of the trace. In the present case, we proceed as follows. We assume 
that the matrices are now operators over some separable Hilbert space 
$\mathcal{H}$. The basis set of this space $|q\rangle , q ~ \epsilon ~ 
\mathbb R$ satisfies the following properties:

\begin{equation}
\label{hbasis}
\int dq~|q\rangle\langle q| = I ~;~ \langle q|q'\rangle = \delta(q - q')
\end{equation}

From which we deduce that the matrix elements:

\begin{equation}
\label{matel}
O(u,v) := \langle u|O|v\rangle
\end{equation}

define completely an operator over this space. The trace of an operator is 
given by:

\begin{equation}
\label{trop}
\int dq~ \langle q|O|q\rangle ~=~ \int dq O(q,q)
\end{equation}

And the product of two operators is given by:

\begin{equation}
\label{prod}
\bigl(O_1O_2\bigr)(p,q) = \langle p|OP|q\rangle = \int dr O_1(p,r)O_2(r,q)
\end{equation}

Where hermitian operator matrix elements satisfy:

\begin{equation}
\label{hermconst}
O_(p,q) = O^*(q,p)
\end{equation}

One can see from the trace of the commutator of a pair of operators:

\begin{equation}
\label{opcom}
Tr [O_1,O_2] = \int dq \langle q|[O_1,O_2]| q \rangle = \int dq dp 
\Bigl(O_1(q,p)O_2(p,q) - O_2(q,p)O_1(p,q)\Bigr)
\end{equation}

that although the integration variables are dummy variables (and hence can 
be interchanged through relabeling), the order of inegration cannot be 
interchaned. If this were the case, the above expression would vanish 
identically. Hence in all that follows, we will meticulously preserve the 
order of integration when writing out operator products, with the order of 
integration is to be read from right to left.\linebreak

\par

Before we can proceed further, we have to address two important issues that 
constrain this model. The first issue is that of the Gauss law constraint 
(\ref{Gauss}), and the second is that our matrices are constrained to be 
hermitian, and in the event that the translational symmetry of this model is broken, of constant trace (we demonstrate further on that this symmetry is  spontaneously broken in the infinite dimensional limit). We will approach the first issue in the usual manner of introducing auxilliary fields. Although one might be tempted to 
tackle the second issue in a similar manner, thus permitting arbitrary field 
variations when applying the variational principle, we choose to account for 
it more directly by only allowing variations that are consistent with 
preserving the hermitian and/or constant trace nature of the matrices. This not only 
does away with what would be a proliferation of auxilliary fields, but turns 
out to be rather easy to implement.\linebreak

\par

Reconsider the constraint:

\begin{eqnarray*}
[\dot{X}_i,X_i] = 0
\end{eqnarray*}

Because of the particular form of our action, we have the luxury of two 
different choices in implimenting this constraint. One could either 
introduce an auxilliary matrix, $\hat{\lambda}$ such that we add the 
following to the action:

\begin{equation}
\label{lmat}
\int dt ~ Tr \bigl(\hat{\lambda} [\dot{X}_i,X_i] \bigr)
\end{equation}

Where requiring the action to be stationary under arbitrary variations of 
the elements of $\hat{\lambda}$ give us the desired constraint. We can 
alternatively introduce an auxilliary function of time, $\lambda$ and add 
the following term to the action:

\begin{equation}
\label{lf}
- \int dt ~ \lambda Tr \bigl([\dot{X}_i,X_i][\dot{X}_j,X_j] \bigr)
\end{equation}

Requiring the action to be stationary under variations of $\lambda(t)$ 
implies:

\begin{equation}
\label{lfeom}
-Tr\bigl([\dot{X}_i,X_i][\dot{X}_j,X_j] \bigr) ~=~ 0
\end{equation}

Since this equation is of the form $TrO^\dag O$, by the positivity of the 
trace norm we can again conclude the desired constraint. It turns out that 
the latter approach is more convenient as it avoids having to solve for the 
constraint explicitly.\linebreak

\par

Turning now to the second issue constraining our model, we see that if our 
variations are to preserve the hermeticity of the matrices, then the 
variations themselves have to be infinitessimal hermitian operators. In 
terms of matrix elements in our Hilbert space this implies:

\begin{equation}
\label{varherm}
\delta X_i(q,p) =  \delta X^*_i(p,q)
\end{equation}

An arbitrary variation that satisfies this constraint is then given by linear combinations of the following:

\begin{equation}
\label{vh2}
\delta_{l,\epsilon,u,v} X_i(q,p) =  \delta_{il} \bigl\{ \epsilon 
\delta(q-u)\delta(p-v) + \epsilon^* \delta(q-v) \delta(p-u)\bigr\}
\end{equation}

Where $\epsilon$ is an arbitrary phase. The meaning of the subscripts on the 
variation should be clear, and will be dropped in the following. If we decide to constrain the model such that the trace of the matrices is to remain fixed (i.e. the variations preserve the trace of the matrices-- a conclusion that as we shall see further on, will be forced on us through the spontaneous breaking of translational symmetry in the infinite N limit) then our variations must also have a vanishing trace. For the case where we consider variations of the form (\ref{vh2}) (we shall consider linear combinations of such variations shortly), this implies:

\begin{equation}
\label{vartrace}
0~ =~ \int dq~ \delta X_i(q,q) ~=~ (\epsilon + 
\epsilon^*)\delta_{il}\delta(u-v)
\end{equation}

This condition is automatically satisfied for variations of off-diagonal 
matrix elements. However for variations of diagonal elements, the phase 
factors would have to be purely imaginary for this expression to vanish. 
This will prove to have striking consequences further on. Note that these 
considerations would also apply to variations of finite dimensional 
matrices, however they do not alter the problem in quite the same way as we 
will see later. Indeed the "central extension" to the model alluded to 
earlier is a phenomenon unique to working with infinite dimensional 
matrices.\linebreak

\par

At this point, it seems reasonable to question when, if ever the trace preserving constraint on our model is applicable since we claim such drastic 
consequences to follow from it. It turns out that this condition is 
redundant for finite dimensional matrices as our equations of motion view 
the trace as a center of mass co-ordinate for the collection of D0 branes 
which moves with a constant velocity, and hence can be set to zero without 
loss of generality. This observation arises from taking the trace of the 
equations of motion-- the trace of $\ddot{X_i}$ gives us the acceleration of 
the center of mass, whereas the trace of the right hand side of the equation, involving only commutators vanishes. This is not the case in infnite dimensions.\linebreak

\par

Now the trace of an unbounded operator is a slippery concept. In 
general it is going to depend on either a regularization scheme or an 
ordering prescription (equivalent to the concept of principal value in 
Riemann integration). To see this, consider the operator $\tau$ whose 
spectrum as we have seen in the integers. The trace of this operator is only 
zero if we count "outwards from zero". That is, $0 + 1 -1 + 2 -2 + 3 - 3... 
= 0$. If we were to evaluate this same sum except now "counting outwards" 
from any other integer, we'd get an infinite number. Thus it would seem to be 
prudent at the very least, to restrict our variations to preserve the trace 
of our operators, whatever they may be, as once we've picked a prescription 
in which to make sense of the trace, arbitrary variations of the matrix 
elements are likely to derail this prescription.\linebreak

\par

However, this argument is not very compelling from a physical perspective, 
and this need not be the case. Intuitively, one would expect that in the 
limit of a truly infinite number of D0 branes, the center of mass 
co-ordinate will cease to be dynamical in the center of mass frame. The action (\ref{act}) is invariant 
under a constant translation of the center of mass coordinate $X_i \to X_i + 
c_i I$, hence it posseses the 3-d translation group as a symmetry. If this 
symmetry is spontaneously broken in the limit of infinite dimensional 
matrices, then we can conclude that the center of mass co-ordinate (hence its velocity, as we chose to work in the centre of mass frame) is 
kinematically superselected to some fixed value and is hence a dynamical 
invariant. We thus have a new constraint on our problem that we had better 
respect when applying the variational principle--  that our variations 
preserve the overall trace of the operators. We explore this possibility in the appendix, in what is essentially an application of the technique of the Coleman-Mermin-Wagner Theorem to our particular model \cite{k16}\cite{k17}\cite{k18}.\linebreak

\par

However we will also see further on that the central extension we are about to derive, can also be derived from another fundamental requirement-- that of Galilean invariance. This invariance arises in the limiting infinite momentum frame in which this matrix model is formulated \cite{bfss}. One finds that requiring this symmetry to be preserved in the limit of infinite dimensional matrices also induces a central extension to the model.\linebreak       

\par

Armed with this knowledge, we can proceed to vary the constrained action. The variation of (\ref{act}) by an arbitrary variation is given by:

\begin{eqnarray}
\label{matact}
\delta S = &T_0&\int dt dq dp 
\Bigl\{\frac{1}{2}\delta\dot{X}_i(q,p)\dot{X}_i(p,q) + 
\frac{1}{2}\dot{X}_i(q,p)\delta\dot{X}_i(p,q) - 
\frac{i\kappa}{3}\epsilon_{ijk}\delta 
X_i(q,p)\bigl([X_j,X_k]\bigr)(p,q)\Bigr\}\\ + &T_0&\int dt dq dp dr 
\Bigl\{\frac{1}{2}\delta X_i(q,r)X_j(r,p)\bigl([X_i,X_j]\bigr)(p,q) - 
\frac{1}{2} X_j(q,r)\delta X_i(r,p)\bigl([X_i,X_j]\bigr)(p,q)\\ &+& 
\frac{1}{2}\bigl([X_i,X_j]\bigr)(q,p)\delta X_i(p,r)X_j(r,q) - 
\frac{1}{2}\bigl([X_i,X_j]\bigr)(q,p) X_j(p,r)\delta X_i(r,q)\\ &-& 
\frac{2i\kappa}{3}\epsilon_{ijk}X_i(q,p)\delta X_j(p,r) X_k(r,q) +  
\frac{2i\kappa}{3}\epsilon_{ijk}X_i(q,p) X_k(p,r)\delta X_j(r,q) \Bigr\}
\end{eqnarray}

And the variation of the constraint term is given by:

\begin{equation}
\label{vartr}
\delta S = -2 \int dt \lambda \Bigl\{ \int dq dp \delta  X_i(q,p) 
\bigl([[\dot{X}_i,[X_j,\dot{X}_j]] \bigr) (p,q) - \delta\dot{X}_i(q,p) 
\bigl( [X_i,[X_j,\dot{X}_j]]\bigr) (p,q) \Bigr\}
\end{equation}

By inserting the explicit form for the variations (\ref{vh2}), and taking care to respect the order of the integrations, extremizing the combined action with the Gauss law constraint term with 
respect to the variation $\delta_{l,\epsilon,u,v}$ implies the following:

\begin{equation}
\label{eomh}
0 = \epsilon\Bigl\{ -\ddot{X_l} + \bigl([X_j,[X_l,X_j]]\bigr) 
-i\kappa\epsilon_{ljk}\bigl([X_j,X_k]\bigr) - 
2\lambda\bigl([\dot{X}_l,[X_j,\dot{X}_j]]\bigr) -2\frac{d}{dt} 
\bigl([X_l,[X_j,\dot{X}_j]]\lambda\bigr) \Bigr\}(u,v) ~+~ \epsilon^*\Bigl\{...\Bigr\}(v,u)
\end{equation}

Where the second expression in the curly brackets differs from the first 
only through an interchange in the variables. As the equations of motion for 
the auxiliary field $\lambda$ imply that the last two terms in the above 
expression vanish, we can drop them right away. Recall that when $u \neq v$, 
the phase factor $\epsilon$ is arbitrary. Hence the above must hold for 
$\epsilon = 1$ and $\epsilon = i$, which is enough to guarantee the 
vanishing of each of the two expressions contained in the larger brackets 
separately, from which we conclude:

\begin{equation}
\label{eomneq}
\ddot{X_l}(u,v) = \bigl([X_j,[X_l,X_j]]\bigr)(u,v) 
-i\kappa\epsilon_{ljk}\bigl([X_j,X_k]\bigr)(u,v) ~;~ u \neq v
\end{equation}

However when $u = v$, were our variations to be restricted by the trace preserving 
condition (\ref{vartrace}), we see that the phase factor for this variation must be purely imaginary. We can see that in this case, (\ref{eomh}) vanishes identically 
and so the principle of least action does not tell us anything. 
The most we can then conclude as our equations of motion is the following:

\begin{equation}
\label{eomgen}
\ddot{X_l}(u,v) = \bigl([X_j,[X_l,X_j]]\bigr)(u,v) 
-i\kappa\epsilon_{ljk}\bigl([X_j,X_k]\bigr)(u,v) + f_l(u)\delta(u-v)
\end{equation}

Where the $f_l(u)$ are some as of yet undetermined functions. Translating 
this expression back into operator language, we have the equations of 
motions:

\begin{equation}
\label{eomop}
\ddot{X_l} = [X_j,[X_l,X_j]] -i\kappa\epsilon_{ljk}[X_j,X_k] + \Delta_l
\end{equation}

Where the $\Delta_l$ are diagonal operators. It is a straight forward exercise to show that we further constrain the $\Delta_l$ to be multiples of the identity, upon considering more general linear combinations of (\ref{vh2}) that satisfy the trace preserving constraint. However a quick and hueristic way of seeing this is to consider an important aspect of the 
action (\ref{act}) that we've neglected up to now-- it's U(N) global 
symmetry. In fact this virtue of the model alone, has generated much 
theoretical interest through the observation that as N goes to infnity, the symmetry group $U(\infty)$ could tend to the group of diffeomorphisms of 
certain 2-dimensional surfaces (see \cite{k13} and references therein). This 
association appears in context of regularizing membrane theory by recasting 
it as a matrix model, where the original worldsheet diffeomorphism 
invariance of the theory translates into a $U(N)$ symmetry \cite{k15}. More 
intriguingly, there are indications \cite{k14} that $U(\infty)$ is much 
bigger than this, and could contain as subgroups the different 
diffeomorphism groups of toplogically distinct manifolds, suggesting that a 
matrix model with this symmetry could be a theory of dynamical D-brane topology 
change. We investigatite this possibility hands on in \cite{k9} (see \cite{topos1} for a similarly spirited investigation).\linebreak

\par

Returning to the problem at hand, we can exploit this $U(\infty)$ invariance 
to deduce that the operators $\Delta_l$ have to be multiples of the 
identity.

\begin{equation}
\label{del}
\Delta_l ~=~ c_l I
\end{equation}

We stress that this particular argument is somewhat hueristic, but the result is nevertheless true from considering arbitrary traceless variations of the action (as dicussed above). Now in the case where $N$ is finite, all one has to do is to take the trace of (\ref{eomop}) to see that the $c_l$ all vanish. However, the fact that the trace of a commutator famously does not vanish when we are dealing with infinite dimensional matrices means that the $c_l$ are now determined by the 
fact that overall, the trace of the right hand side of (\ref{eomop}) has to vanish by virtue of the fact that the trace of $X_l$ has to vanish. It is this fact that will permit us to look for new solutions to this matrix model in the limit of infnite dimensional matrices. For starters, we can take the $c_l$ to parametrize new solutions which may be central extensions of the usual algebras taken to define non-commutative geometries. However the most obvious new solution permitted is the non commuting plane-- the equations of motion (\ref{lc1})(\ref{lc2})(\ref{lc3}) now become:

\begin{equation}
\label{lcc1}
\ddot{X}_+ ~= ~\frac{1}{2}[X_+, [X_+,X_-] ] ~+ ~[X_3, [X_+,X_3] + 2\kappa 
X_+ ] ~+~ c_+ I
\end{equation}

\begin{equation}
\label{lcc2}
\ddot{X}_- ~= ~\frac{1}{2}[X_-, [X_-,X_+] ] ~+ ~[X_3, [X_-,X_3] - 2\kappa 
X_- ] ~+~ c_- I
\end{equation}

\begin{equation}
\label{lcc3}
\ddot{X}_3 ~= ~\frac{1}{2}[X_+, [X_3,X_-] ] ~+ ~\frac{1}{2}[X_-, [X_3,X_+] ] 
~+ ~\kappa [X_+,X_-] + c_3 I
\end{equation}

From which we see that in the static case, for $c_\pm = 0$, $c_3 = \beta$:

\begin{equation}
\label{ncpl}
X_3 = \lambda I~,~ X_+ = a^\dag~,~X_- = a~;~ [a,a^\dag] = \frac{\beta}{\kappa}
\end{equation}

also solves our equations of motion (the case where all the $c_i$ are zero would give us the fuzzy cylinder and the fuzzy sphere solutions). Here, $\lambda$ is a constant describing the displacement of the fuzzy plane along the z-direction.  Hence in the limit of an infnite 
collection of D0 branes in a RR background, a D2 brane forms with the 
structure of the non-commutative plane in addition to the fuzzy cylinder and fuzzy two sphere solutions. The non-commutativity of course 
being parametrized by the ratio of $c_3$ over the magnitude of the RR field strength given by  $\kappa$. Interestingly, unlike the cases hitherto studied, the degree of non-commutativity dependends inversely on the background 4-form field strength. It should be clear that the $c_i$ now serve to parametrize different matrix models, quite like the central charge of a given conformal field theory. We can of course pick certain values for them and look for new solutions, but it would be desirable to understand the $c_i$ in terms of more fundamental considerations. Such an understanding is readily available and we investigate this in the next section. However, to round out this part of the discussion, we study the energetics of these infinite dimensional solutions. It is straightforward matter to find the energy density $\epsilon$ ( = E/Tr I) of the fuzzy cylinder:

\begin{equation}
\label{ecyl}
\epsilon_{cyl} = T_0 \frac{2}{3}\kappa^2\rho^2
\end{equation}    

And the energy density of the fuzzy plane:

\begin{equation}
\epsilon_{pl} = T_0\frac{\beta^2}{8\kappa^2}
\label{encpl}
\end{equation}

Where we should point out that the form of the Hamiltonian is no different in the infinite dimensional case. That the central extensions do not alter the energy of our solutions will be justified in the next section. As these are positive energies, one cannot interpret these solutions as bound states just yet, as one would have to study the local energy landscape around these solutions to test their (meta) stability. Such a study would have taken us far afield and we leave this to a follow up report.
Since this concludes the main thrust of this report, we feel that a brief summary of our treatment is in order at this time. \linebreak

\par

In studying the matrix model defined by (\ref{act}), we realised through 
naive manipulation of the static solutions for finit N that in addition to 
admitting fuzzy 2-sphere geometries as fundamental solutions, it appeared 
that fuzzy cylinders could also be solutions. Such solutions would 
neccesarily have been infinite dimensional as the lie algebra defining the 
fuzzy cylinder is that of the euclidean group in 2-dimensions, and as such 
is a non semi-simple algebra. We carefully considered the transition to 
infnite dimensional matrices in our model, and not only confirmed that the 
cylinder was indeed a solution, but discovered the fact that in general this 
model generically has central extensions. These central extensions are facilitated through the spontaneous symmetry breaking of translational invariance of this model in the limit of infinite dimensional matrices, and as we are about to see, through requiring that Galilean invariance be preserved in this limit. These extensions parametrize 
different models, and for special cases we are likely to find new fuzzy 
geometries as solutions. We considered a special central extension that 
permitted the non-commuting plane as a solution, thus further confirming the 
richness of physics contained in this matrix model. We hope to use the 
results here in a future report to investigate the rather pregnant possibility of dynamical topology change \cite{k9} that this model seems to suggest as possible.

\section{physical interpretation of the central extension}

We show in the appendix how the central extensions just uncovered, owe their origins to the spontaneous breaking of translational symmetry of the center of mass degree of freedom in the limit of infinite dimensional matrices. It should be clear that these central extensions can also be obtained by imposing the constraint that the center of mass degree of freedom be fixed through the method of Lagrange multipliers. Such a constraint can be enforced through the addition of the term:

\begin{eqnarray*}
S_{constr} &=& \int dt~\sigma_i\Bigl(\frac{1}{Tr I}TrX_i - \lambda_i \Bigr)\\
&=& \int dt~ \frac{1}{Tr I}Tr \sigma_i(X_i - \lambda_iI)
\end{eqnarray*}    

where the $\sigma_i$ are the functions of time which act as Lagrange multipliers, and the $\lambda_i$ describe values that the c.o.m coordinates are constrained to. Such an addition to the action will add the following "central" term to our equations of motion:

\begin{equation}
\label{trcom}
\ddot{X}_i = [X_j,[X_i,X_j]] -i\epsilon_{ijk}[X_j,X_k] + c_i I ~ ; ~c_i = \sigma_i/Tr I 
\end{equation} 

Where we see that the $c_i$ we derived earlier are none other than the appropriate Lagrange multipliers rescaled. We expect the $c_i$ to be non zero despite the factor of $Tr I$ in the definition, as the Lagrange multipliers are also expected to scale proportional to $Tr I$. This observation arises from the interpretation of the Lagrange multipliers as the appropriate force term enforcing the constraint, which we expect to scale with the number of degrees of freedom in the problem \cite{gold}. We should note here that this understanding of the $c_i$ allows us to conclude that it would not contribute to the Hamiltonian of our model, as its constribution would vanish through the equations of motion for the lagrange multipliers. We should also note that inspite of trying from the outset to avoid introducing unneccesary auxiliary fields, we see that for this particular constraint, it could not be avoided. This realisation arises when searching for the physical interpretation of the $c_i$.\linebreak

\par

Having made this connection, one might also wonder what the effect would be of imposing as a constraint, something that one might ordinarily take forgranted from the Galilean invariance inherent in the model. Starting from the begining, let us add instead, for example the following term to our action: 

\begin{eqnarray*}
S_{constr} &=& \int dt~ \sigma_i \Bigl( \frac{1}{Tr I} Tr \dot{X}_i - \omega_i\Bigr)\\
&=& \int dt~ \frac{1}{Tr I} Tr \sigma_i(\dot{X}_i - \omega_iI)
\end{eqnarray*}  

Such a term implies that the center of mass momentum is a constant. One would expect such an introduction to be redundant, by Galilean invariance. However examining the equations of motion without this term introduced into the action:

\begin{equation}
\label{iuirery}
\ddot{X}_i = [X_j,[X_i,X_j]] -i\epsilon_{ijk}[X_j,X_k]
\end{equation}  

We see that in the case of finite dimensional matrices, Galilean invariance is trivially implied for the center of mass momentum. However since it cannot be taken forgranted that commutators have a vanishing trace in infinite dimensions, we see that one might have a violation of Galilean invariance unless a compensating term appears in the equations of motion to account for this peculiar property of infinite dimensional matrices. Such a term is readily provided by the constraint term introduced above:

\begin{equation}
\label{trcom2}
\ddot{X}_i = [X_j,[X_i,X_j]] -i\epsilon_{ijk}[X_j,X_k] + c_i I~:~ c_i = -\dot{\omega}_i/ Tr I
\end{equation}  

Where the only difference with (\ref{trcom}) is in the definition of the $c_i$. Thus we see that the central extensions to this matrix model can be deduced from seemingly disparate, but equally well grounded physical considerations.

\section{more solutions}

In addition to the fuzzy plane solutions (\ref{ncpl}), we find that the same central extension (parametrized by $c_\pm = 0$, $c_3 = \beta$):

\begin{eqnarray*}
\ddot{X}_+ ~&=& ~\frac{1}{2}[X_+, [X_+,X_-] ] ~+ ~[X_3, [X_+,X_3] + 2\kappa 
X_+ ]\\
\ddot{X}_- ~&=& ~\frac{1}{2}[X_-, [X_-,X_+] ] ~+ ~[X_3, [X_-,X_3] - 2\kappa 
X_- ]\\
\ddot{X}_3 ~&=& ~\frac{1}{2}[X_+, [X_3,X_-] ] ~+ ~\frac{1}{2}[X_-, [X_3,X_+] ] ~+ ~\kappa [X_+,X_-] + \beta I
\end{eqnarray*}

is also solved by the algebra: 

\begin{equation}
\label{warp}
[X_3,X_\pm] = \pm 2\kappa X_\pm ~,~ [X_+,X_-] = \frac{\beta}{\kappa} I
\end{equation}

One can easily furnish a representation of this algebra through the identifications:

\begin{eqnarray}
X_+ &=& a~,~X_- = a^\dag~~;~~[a,a^\dag] = \frac{\beta}{\kappa}\\ 
X_3 &=& \frac{2\kappa^2}{\beta}a^\dag a
\end{eqnarray}

Where we have reversed the identifications in (\ref{ncpl}) with $X_+ = a$ and $X_- = a^\dag$. Although from (\ref{warp}) one might view this as a deformation of the algebra of the cylinder, it has a clearer geometric interpretation as a deformation of the fuzzy plane (\ref{ncpl}) found earlier. One obvious reason for this is that the usual representation of the oscillator algebra is spanned by state vectors indexed by the non-negative integers, whereas the cylinder algebra is spanned by state vectors indexed by all the integers \footnote{See discussion surrounding (\ref{shift})}. Furthermore (\ref{warp}) can be interpreted easily in terms of the fuzzy plane, where instead of being restricted to a surface of constant $X_3$, it parabolically "warps" the further "out" you go (recall that $X_+X_-$ can be interpreted as a radial co-ordinate, where the identification is exact in the limit of commuting $X_+$ and $X_-$). Since $X_3$ is quantized in integer steps of $\beta/\kappa$, it would seem that this solution has warped the fuzzy plane into a parabola-like fuzzy configuration. \linebreak

\section{closing remarks}

We conclude our report at this point, with the hope that the new solutions we have uncovered serve to demonstrate several (of the likely many possible) novel features of this model in the limit of infinite dimensional matrices. We feel that this may yet be a shadow of things to come, as previously in the discussion we raised the possibility that this model may be a candidate model of topological dynamics in the large N limit. We will investigate this possibility in a follow up report \cite{k9}, where transitions between the fuzzy cylinder and the (multiple) fuzzy sphere solutions will be investigated. Independent of this however, we feel that an equally fruitful avenue for future work would be to study the string theoretical interpretation of these results, which we have so far not attempted since we have deliberately chosen to study this model from the perspective of a non-commutative geometer. Certainly such a study is likely to uncover yet more of the richness of this model and perhaps shed further light on the some of questions that motivated this study, such as that of topological and geometrical dynamics in matrix models.

\section{Acknowledgements}

The author wishes to express deep gratitude to the Theory group (and Sandip Trivedi in particular) at the Tata Institute for Fundamental Research in Mumbai, for their hospitality during the time in which this work was undertaken. The support and encouragment they provided, along with that of Robert Brandenberger, Rudra Jena and many others during what were rather difficult circumstances is greatfuly acknowledged. We wish to thank Gautam Mandal for many valuable discussions and inspiration, and Pei Ming-Ho for valuable comments on the manuscript.

\begin{appendix}
\section{Spontaneous Symmetry Breaking as N $\to \infty$}

Consider the action for the center of mass 
degrees of freedom, $x_i = \frac{Tr X_i}{Tr I}$:

\begin{equation}
\label{comact}
S = T_0 \int  dt ~ Tr I ~ \dot{x}_i\dot{x}_i
\end{equation}

Where we have used the relation that the kinetic energy $ = p_{com}^2/2m$. Now consider the quantum mechanical transition amplitude bewteen two states 
that differ by some center of mass displacement:

\begin{equation}
\label{transamp}
\langle \vec{x}(T) | \vec{x}(0) \rangle ~=~ \int D\vec{c} ~ e^{-T_0 \int^T_0 
  dt ~ \dot{x}_i\dot{x}_i Tr I}
\end{equation}

Where we are functionally integrating over all histories that begin at 
$\vec{x}(0)$ and end at $\vec{x}(T)$ with $\vec{d} = \vec{x}(T) - 
\vec{x}(0)$. Upon integration, any such amplitude will contain a prefactor 
coming from the classical path interpolating between these two states:

\begin{equation}
\label{classpath}
\vec{x}(t) = \vec{x}(0) + \vec{x}\frac{t}{T}
\end{equation}

Which will take the form (since $Tr I = N$)

\begin{equation}
\label{ampi}
\langle \vec{x}(T) | \vec{x}(0) \rangle ~=~ e^{-T_0 N \frac{d_id_i}{T}} 
\langle ... \rangle
\end{equation}

Where the remainder of the amplitude comes from integrating the fluctuations 
around this solution. From the prefactor, we can immediately see that this 
amplitude is vanishingly small in the limit $N \to \infty$. The only way 
this amplitude can have a finite prefactor is if the time taken for this 
interpolation, $T$ becomes very large or if the displacement $\vec{d}$ 
becomes vanishingly small. This implies that the center of mass coordinate 
has to move infinitessimally slowly over very small distances. This is the 
same situation as if the center of mass were fixed. Hence we can conclude 
that the translation symmetry of (\ref{act}) is spontaneously broken in the 
limit of infinite dimensional matrices. The displacement of the center of 
mass coordinate of our gas of D0 branes is thus superselected to be some fixed value and in applying the variational principle, we should respect this 
superselection rule. \linebreak 

\par

It is instructive to make a comparison with the equivalent result in field theory (which illustrates the general conclusion that there are no Goldstone bosons in 1+1 dimensions \cite{k18}). Consider a free scalar field theory in 1+d dimensions:

\begin{equation}     
\label{sft}
S = \frac{1}{2} \int dtd^dx~ \partial_\mu \phi \partial^\mu \phi
\end{equation}

In the absence of a potential energy term, the value of the field serves as a modulus labelling different vacua. Consider an instanton transition between two different vacua given by the solution:

\begin{equation}
\label{instanton}
\phi(t,\vec{x}) = \phi_0 + \frac{t}{t_f}(\phi_f - \phi_0)
\end{equation}

where $t_f$ is the time taken to complete the transition. Evaluating the action functional of the classical path interpolating between the two vacua gives us:

\begin{equation}
S_{cl} = \frac{1}{2}\frac{(\phi_f - \phi_0)^2}{t_f}V
\end{equation}   

Thus if we begin by studying this theory in a d-dimensional box of length L, and take the limit $t_f \to \infty$, $L \to \infty$ isotropicaly, we see that the classical prefactor multiplying the quantum transition amplitude becomes:

\begin{equation}
e^{-\frac{1}{2}(\phi_f - \phi_0)^2L^{d-1}}
\end{equation} 

From which we see that if $d \geq 2$, this transition amplitude vanishes and we have spontaneous symmetry breaking of the translation symmetry in $\phi$. However when $d = 1$, this amplitude remains finite and hence transitions between the different vacua are generically possible, hence this symmetry is not spontaneously broken (no goldstone bosons in 2-d).

\end{appendix}

\end{flushleft}

\end{document}